\documentclass[aps,prd,nofootinbib,twocolumn,floatfix,superscriptaddress,eqsecnum]{revtex4}
\usepackage{dcolumn}
\usepackage{times,mathptm}
\usepackage{subfigure}
\usepackage{graphicx,psfrag,pifont,color,setspace}
\usepackage{amsfonts,amsthm,bm,bbm,amsmath,amssymb,marvosym} 
\usepackage[automark]{scrpage2}
\newcommand{\beq}{\begin{equation}}
\newcommand{\eeq}{\end{equation}}
\newcommand{\bea}{\begin{eqnarray}}
\newcommand{\eea}{\end{eqnarray}}

\renewcommand{\l}{\lambda}

\renewcommand{\b}{\beta}

\renewcommand{\r}{\rho}
\newcommand{\psib}{\overline{\psi}}
\newcommand{\bx}{{\mathbf{x}}}
\newcommand{\by}{{\mathbf{y}}}

\newcommand{\s}{\sigma}

\newcommand{\D}{{\Delta}}

\newcommand{\oh}{{\textstyle{\frac{1}{2}}}}

\newcommand{\dg}{\dagger}
\newcommand{\non}{\nonumber}

\newcommand{\rf}[1]{(\ref{#1})}

\begin{document}

\title{Constituent Gluon Content of the Static Quark-Antiquark State
in Coulomb Gauge}

\author{J. Greensite}
\affiliation{Physics and Astronomy Dept., San Francisco State
University, San Francisco, CA~94132, USA}

\author{{\v S}. Olejn\'{\i}k}
\affiliation{Institute of Physics, Slovak Academy
of Sciences, SK--845 11 Bratislava, Slovakia}

\date{\today}
\begin{abstract}
     Motivated by the gluon-chain model of flux tube formation, we compute and diagonalize the transfer 
matrix in lattice SU(2) gauge theory for states containing heavy static quark-antiquark sources, with separations up
to one fermi.  The elements of the transfer matrix are calculated by variational Monte Carlo methods, in a basis of 
states obtained by acting on the vacuum state with zero, one, and two-gluon operators in Coulomb gauge.   The 
color Coulomb potential is obtained from the zero gluon to zero gluon element of the transfer matrix, and it is well-known
that while this potential is asymptotically linear, it has a slope which is two to three times larger than the standard asymptotic string tension.  
We show that the addition of one and two gluon states results in a potential which is still linear, but  the disagreement with the
standard asymptotic string tension is reduced to 38\% at the largest lattice coupling we have studied.

\end{abstract}

\maketitle

\section{\label{intro}Introduction}

     It is well known that the color Coulomb potential in non-abelian pure gauge theories is asymptotically linear, but that the corresponding string tension $\s_{coul}$ is greater than the standard string tension by anywhere from 100-200\%, depending on the lattice coupling and the gauge group \cite{GO,japan}.  This discrepancy should not be too surprising, since the color Coulomb potential is only an upper bound on the static quark potential \cite{Dan}, but the more relevant fact is that the Coulombic force arises from (dressed) one-gluon exchange between static sources.   A serious criticism of one-gluon exchange models of the confining potential, whether in Coulomb or covariant gauges, is that they provide no explanation whatever for the collimation of color electric fields into flux tubes, and, as a consequence, no obvious reason for the absence of long-range dipole forces.   Since long-range dipole forces are incompatible with a massive spectrum, a theory of confinement via dressed one-gluon exchange is in danger of inconsistency, as well as conflicting with the numerical evidence for color-electric flux tubes.
 
\begin{figure}[h!]
\centerline{\scalebox{0.35}{\includegraphics{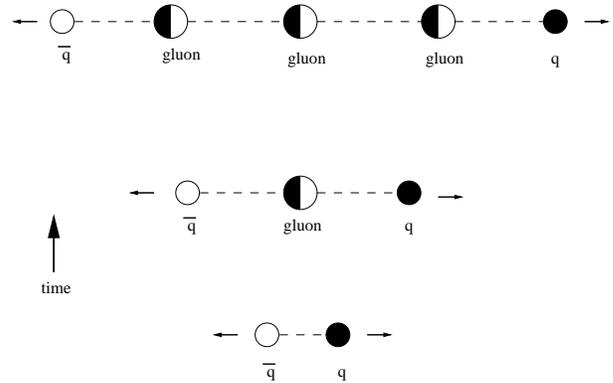}}}
\caption{The gluon chain model.  As a quark-antiquark pair moves apart, they pull out a chain of constituent gluons between them.  Open/filled shadings of the constituent gluons represent the two matrix indices of the $A^{ab}_k$ operator.  Dashed lines indicate both repeated matrix indices, and 
nearest-neighbor Coulomb interactions.}
\label{chain}
\end{figure}

     There is, however, a picture of how flux tubes can form in Coulomb gauge, in which strong one-particle exchange forces are an important ingredient. This is the ``gluon-chain" model, put forward by Thorn and one of the present authors  \cite{GT}.  The idea is that as a heavy quark and and antiquark separate, they pull out a sequence of constituent gluons between them, as illustrated in Fig.\ \ref{chain}.  The constituent gluons are bound together by Coulombic nearest-neighbor interactions, and the ensemble of gluons resembles a discretized string.  In the original proposal it was supposed that the Coulombic force at some distance scale rises faster than linearly, and that the role of the constituent gluons is to prevent a corresponding faster-than-linear rise in the static quark potential, by effectively placing an upper limit on color charge separation.   We now know that the Coulomb potential is itself linear at large scales.  So the role of constituent gluons must be to somehow reduce the magnitude of the static quark potential, without at the same time destroying the linearity property.   That such an effect might take place was shown analytically by Krupinski and Szczepaniak \cite{Adam}, in the context of a particular proposal for the form of the Coulomb gauge vacuum state.  It is desirable, however, to see if constituent gluons can have this conjectured effect without making any assumptions about the form of the vacuum.  That goal, at present, can only be achieved numerically, via lattice Monte Carlo simulations.

    There has been only one attempt \cite{evidence}, carried out twenty years ago, to study the constituent gluon content of the QCD flux tube in Coulomb gauge by numerical simulation, and thereby test the gluon chain model.   Our intention in the present article is to greatly improve on that earlier effort, and to quantify the reduction in the energy of a static quark-antiquark state which can be achieved with a handful of constituent gluon operators.
     
\section{\label{sec2}The transfer matrix in a variational basis} 

    The Euclidean-time evolution operator in lattice gauge theory, in some physical gauge such as Coulomb gauge, is the transfer matrix
\beq
      {\cal T}=\exp[-Ha]
\eeq
where $a$ is the lattice spacing and $H$ the Hamiltonian. It is useful to consider the rescaled operator
\bea
      T &\equiv& {  {\cal T} \over \langle \Psi_0| {\cal T}|\Psi_0\rangle }
\non \\
         &=& \exp[-(H-E_0)a]
\eea
where $\Psi_0,~E_0,$ are the ground state and vacuum energy, respectively.   To compute the static quark potential of a quark-antiquark pair separated by a distance $R$, one would ideally diagonalize the transfer matrix in the infinite-dimensional subspace of states which contain a single massive quark, and a single massive antiquark,  located at sites $\bx$ and $\by$ with $R=|\bx-\by|$.   The minimal energy eigenstate of the transfer matrix, in this subspace, is the state with the largest eigenvalue $\l_{max}$ of $T$, and the static quark potential, in lattice units, is given by
\beq
            V(R) =   - \log(\l_{max})
\eeq
(from here on we will work in lattice units).
If we consider any two states $|a\rangle$ and $|b\rangle$ which are obtained by acting on the vacuum state with operators $Q_a$ and $Q_b$,
respectively, then 
\beq
     \langle a |T|b \rangle = \langle Q^\dg_a(t+1) Q_b(t) \rangle
\label{TQQ}
\eeq
where the notation $Q(t)$ indicates that the operator is to be evaluated using link variables on a hypersurface of fixed time $t$.  For given
operators $Q_{a,b}$, the rhs of eq.\ \rf{TQQ} can be evaluated by lattice Monte Carlo simulation.

    If the quark and antiquark are not too far apart, then according to the gluon-chain picture it should be possible to express the minimal energy state
in terms of the vacuum state $\Psi_0$, and a handful of operators operators acting on that state.  These operators create a massive antiquark at lattice site $\bx$, a massive quark at site $\by$, and a small number of constituent gluons.  Let us denote the gluonic operators by $Q_k$, deferring their actual construction to the next section, with corresponding states
\beq
              |k \rangle =  \psib^a(\bx) Q_k^{ab}  \psi^b(\by) |\Psi_0 \rangle  ~~~~k=1,2,..,M
\eeq
The $\{Q_k\}$ are functionals of the lattice gauge field, and depend on $\bx,~\by,$ and some number of variational parameters. In our construction, described below in section \ref{sec3}, we use a single variational parameter, denoted $\rho$. 
One can then compute, by lattice Monte Carlo simulation, the quantities
\bea
       O_{mn} &=& \langle m | n \rangle 
\non \\ &=& \langle \oh \mbox{Tr} [Q^\dg_m(t) Q_n(t) ] \rangle
\non \\
       t_{mn}  &=&  \langle m |T| n \rangle 
\non \\
  &=&  \langle \oh \mbox{Tr} [Q^\dg_m(t+1) U^\dg_0(\bx_0,t) Q_n(t) U_0(\bx_L,t) ] \rangle
\label{Nt}
\eea 

  We cannot diagonalize the transfer matrix $T$ in the basis $\{ |k \rangle \}$, because these states are not orthogonal, in general, nor are
they normalized.  So the next thing to do is to construct a new set of orthogonal states $\{ | k' \rangle \}$ from the non-orthogonal states 
by the Gram-Schmidt orthogonalization procedure
\bea
|k'\rangle  &=&  |k\rangle - \sum_{j=1}^{k-1} {\langle j'|k\rangle \over \langle j' |j'\rangle } |j'\rangle 
\non \\
  &=& \sum_{j=1}^{k} D_{kj}  |  j \rangle 
\eea
Defining
\bea
N_j &=& \langle j' | j' \rangle
\non \\
&=& \sum_m \sum_n D_{jm} D_{jn} \langle m |  n \rangle
\eea
we have
\bea
|k'\rangle  &=&  |k\rangle - \sum_{j=1}^{k-1}{1\over N_j} \sum_m D_{jm} \langle m |  k \rangle 
         \sum_n D_{jn} |n \rangle
\non \\
&=& |k\rangle - \sum_{n=1}^{k-1}\left[ \sum_{j=1}^{k-1} {D_{jn}\over N_j}  \sum_{m=1}^{j} D_{jm} \langle m |  k \rangle 
\right]  |n \rangle
\eea
The $D_{mn}$ coefficients can be determined iteratively by the following set of relations:
\beq
            D_{kn} = \left\{ \begin{array}{cl}  - \sum_{j=1}^{k-1}  N^{-1}_j D_{jn}  \sum_{m=1}^{j} D_{jm} \langle m |  k \rangle  & n<k \cr
                           1 &  n=k \cr 
                           0  &  n>k  \end{array} \right.
\eeq    
Normalize,
\bea
         |\phi_k \rangle &=&  {1\over \sqrt{N_k}} |k'\rangle
\non \\
&=&  {1\over \sqrt{N_k}}  \sum_{m=1}^k D_{km} |m \rangle
\eea
and calculate the $M\times M$ matrix 
\bea
          T_{ij} &=& \langle \phi_i | T | \phi_j \rangle
\non \\
&=& \sum_{n=1}^i \sum_{m=1}^j   {1\over \sqrt{N_i}}  {1\over \sqrt{N_j}} D_{in} D_{jm} \langle n|T|m\rangle
\non \\
&=&  {1\over \sqrt{N_i N_j}} \sum_{n=1}^i \sum_{m=1}^j   D_{in} D_{jm} t_{nm}
\eea 
Note that this matrix is derived from the quantities $O_{mn}$  
and $t_{mn}$ of eq.\ \rf{Nt}, both of which are calculated by lattice Monte Carlo.     

    The final step is to extract the largest eigenvalue $\l_{max}$ of the finite matrix $T_{ij}$.  Since this is a variational approach, 
the computation has to be repeated for a variety of values of the variational parameter $\rho$, at each quark-antiquark separation $R$.   
The value of $\r$ which minimizes $-\log(\l_{max})$ gives the best variational estimate
\beq
V_{chain}(R) = -\log(\l_{max})
\eeq
for the static quark potential.
We will refer to this variational result as the ``gluon-chain potential", although in practice the variational state will be a superposition of states containing at most two constituent gluons. 
    
\section{\label{sec3}Smeared one and two-gluon operators}

    In order to actually carry out the procedure described above, we must specify a set of operators $\{Q_k^{ab}\}$ which, operating
on the Coulomb gauge vacuum, generate a set of variational trial states $\{|k\rangle\}$.   

    A gluon chain consists of a number of constituent gluons lying between the heavy sources, with the ordering of the gluons in their color-indices correlating with the spatial positions of gluons between the static sources.   Suppose, e.g., that the static quark-antiquark sources are located on the $x$-axis at locations $x=0$ and $x=R$.  Constituent gluons which are located well outside this interval in their $x$-coordinates, or which have large transverse displacements away from the $x$-axis, will be costly in terms of Coulombic interaction energy (recall that this energy itself increases linearly with color charge separation). On the other hand, highly localized constituent gluons are costly in terms of kinetic energy.  As usual, the variational procedure attempts to find an optimal compromise between kinetic energy, which favors spatial delocalization, and interaction energy, which in this case favors small transverse displacement from the line joining the quark sources.   Delocalization in the $x$-direction is achieved by a superposition of gluon operators (in the $Q_k$) at different locations along the $x\in [0,R]$ interval.\footnote{In the two-gluon states described below we have enlarged this interval slightly, allowing for $x$ coordinates up to two lattice spacings outside the interval.}    Delocalization in the transverse directions can be obtained by constructing $A$-field operators on the lattice, in which the high-frequency components of the $A$-field, in the transverse directions, are gaussian suppressed.   

     The delocalized, or ``transverse-smoothed'', operators $A^c_i(\bx,t,j)$ are constructed in the following way:  The $L^4$ lattice is fixed to Coulomb gauge by standard methods (simulated annealing + over-relaxation),  and the usual lattice $A$-field variable, in the SU(2) gauge group, is defined in terms of the link variables via
\bea
          A_k(\bx,t) &=& {1\over 2i}\Bigl( U_k(\bx,t) - U_k^\dg(\bx,t) \Bigr)
\non \\
                      &=& A^c_k(\bx,t) {\s_c \over 2} 
\eea
For each lattice we loop through time-slices $t$,  space components $i=1,2,3$, color components $c=1,2,3$, and an additional direction 
$j=1,2,3$, which will be associated with the direction of a line through the quark and antiquark.  At each $i,j,c,t$, and denoting $\bx=(n_1,n_2,n_3)$, (with indices $n_i$ running from $0$ to $L-1$) define the three-dimensional array
\beq
          a(n_1,n_2,n_3) = A_i^c(\bx,t) 
\eeq 
Denote the finite Fourier transform of this array by $\widetilde{a}(n_1,n_2,n_3)$, with indices running over the same range of $0$ to $L-1$, and define
\beq
          f_n = \left\{ \begin{array}{cl}
                   n  & ~~~0\le n \le {L\over 2} -1 \cr
                 n-L & ~~~{L\over 2} \le n \le L-1 \end{array} \right.
\eeq
The wavenumber corresponding to index $n_i$ is $k_i = 2\pi f_{n_i}/L$.  Transverse-smoothing is achieved by making an exponential 
suppression of the large wavenumber modes in the directions $l,m$ which are transverse to the direction $j$, and transforming the result back to position space.   The prescription is to modify the Fourier-transformed array elements by the replacement
\beq
          \widetilde{a}(n_1,n_2,n_3) \rightarrow {1\over L^3} \exp\Bigl[-\rho(f_{n_l}^2 + f_{n_m}^2)\Bigl] \widetilde{a}(n_1,n_2,n_3) 
\eeq
where $\rho$ is a variational parameter, followed by an inverse Fourier transform back to position space.   Then
\beq
         A_i^c(\bx,t,j) = a(n_1,n_2,n_3)
\eeq
is the field variable $A^c_i$ on a time-slice $t$, with suppression of the large wavenumber components in the
$l,m$-directions transverse to $j$.  From this, we define the transverse-smoothed matrix-valued field variable
\beq
        A_i(\bx,t,j) = A_i^c(\bx,t,j) {\s_c \over 2}
\eeq
In the same way, define
\beq
             B_i(\bx,t) = 1 - \oh \mbox{Tr}[U_i(\bx,t)]
 \eeq
Then, just as with the $A^c_i(\bx,t)$, suppress high-wavenumber components in directions transverse to some direction $j$, as described above.  Denote the resulting field as $B_i(\bx,t,j)$.  It is also convenient to define for $i\ne j$, the average
\bea
          \overline{A}_i(\bx,t,j) &=& \oh (A_i(\bx,t,j) + A_i(\bx - \mathbf{e}_i,t,j) 
\non \\
          \overline{B}_i(\bx,t,j) &=& \oh (B_i(\bx,t,j) + B_i(\bx - \mathbf{e}_i,t,j) 
\eea
derived from two links in the $i$ direction which touch the line, in the $j$ direction, running from quark to antiquark.

    Having obtained the transverse-smoothed field variables $A_i(\bx,t,j),~B_i(\bx,t,j)$, the next step is to define the six operators $\{Q_k\}$
from which we obtain six states
\beq
             |k \rangle  =  Q_k |0\rangle ~~~(k=1-6)
\eeq
which are used to build the trial gluon-chain state.  Let the antiquark and quark charges lie at lattice sites 
$\bx_0$ and $\bx_R = \bx_0 +  R \mathbf{e}_j$, respectively, where $\mathbf{e}_j$ is a unit vector in the $j$-direction.\footnote{Of course, in evaluating the transfer matrix elements $T_{nm}$ by lattice Monte Carlo simulation, we average over all $\bx_0$ and $j=1,2,3$.}  Then we choose
\bea
Q_1(t) &=& \mathbbm{1}_2
\non \\
Q_2(t) &=& \sum_{n=0}^{R-1} A_j(\bx_0+n \mathbf{e}_j,t,j)
\non \\  
Q_3(t) &=& \sum_{n=-2}^{R+1} ~ \sum_{n'=n}^{R+1} A_j(\bx_0+n \mathbf{e}_j,t,j) A_j(\bx_0+n' \mathbf{e}_j,t,j)
\non \\
Q_4(t) &=& \sum_{n=-2}^{R+2} ~ \sum_{n'=n}^{R+2} ~ \sum_{i\ne j}
         \overline{A}_i(\bx_0+n \mathbf{e}_j,t,j) \overline{A}_i(\bx_0+n' \mathbf{e}_j,t,j)
\non \\
 Q_5(t) &=& \sum_{n=0}^{R-1} B_j(\bx_0+n \mathbf{e}_j,t,1) \mathbbm{1}_2
\non \\
 Q_6(t) &=& \sum_{n=0}^{R-1} \sum_{i\ne j} \overline{B}_i(\bx_0+n \mathbf{e}_j,t,j) \mathbbm{1}_2
\eea
$Q_1$ is the zero constituent gluon operator,   $Q_2$ is the one gluon operator (one power of $A$), and the $Q_{3-6}$ are 
two-gluon operators, containing two powers of the $A$-field.  The corresponding set of states $\{|k\rangle\}$ are not orthogonal, but this is taken care of by the Gram-Schmidt orthogonalization procedure described in the last section.   

     The above choice of operators $\{Q_k\}$ is dictated by simplicity (only one variational parameter), and a certain amount of trial and error.  It is certainly possible to invent operators creating more general and sophisticated trial states, at the cost of additional variational parameters.  
     
\section{\label{results}Results}

     We have carried out the calculation outlined in section 2, with the operators $Q_k$ described in section 3, for SU(2) lattice gauge theory
at coupling $\b=2.2$ on a $12^4$ lattice volume, $\b=2.3$ on a $16^4$ lattice volume, and $\b=2.4$ on a $22^4$ lattice volume. 
The operators $Q_k$ introduced in the previous section contain a variational parameter $\r$, and we have calculated matrix elements $T_{ij}$ and 
$V_{chain}(R)$ for each $R$, at twelve values of $\rho$ 
\beq
       \rho_n = (n-1) \D \rho ~~~~~ (1\le n \le 12)
\eeq
with $\D \rho = 0.025$ at $\b=2.2$, and $\D \rho=0.02$ at $\b=2.3,2.4$.  The choice of $n$ which minimizes $V_{chain}(R)$ of course
depends on both $\b$ and the quark separation. For example, the best choice at $\b=2.2$ and quark separation $R=1$ was $n=2$.
At $\b=2.4$ and $R=9$, the optimal value was $n=8$.  The results reported below are those obtained at the optimal value of $\rho$
in each case.

    One matrix element of $T$ which does not depend on the variational parameter is the zero-gluon to  zero-gluon matrix element
\beq
          T_{11} = \langle \phi_1 | T| \phi_1 \rangle =  \langle 1 | T| 1 \rangle
\eeq
The color Coulomb potential reported in refs.\ \cite{GO,japan} is derived from this matrix element, i.e.
\beq
           V_{coul}(R) = -\log(T_{11})
\eeq 
For purposes of comparison, we have also computed the usual static quark potential $V_{true}(R)$, which is the minimal possible field energy of a static quark-antiquark system, by the standard method of computing timelike Wilson loops with ``fat" spacelike links, and looking for a plateau in the lattice logarithmic time derivative, as described in, e.g., ref.\ \cite{Bali}.   

\begin{figure}[h!]
\subfigure[]  
{   
 \label{pot22}
 \includegraphics[scale=0.7]{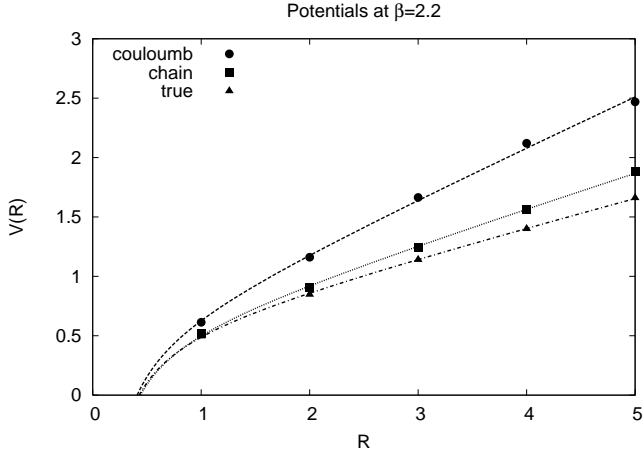}
}
\subfigure[]  
{   
 \label{pot23}
 \includegraphics[scale=0.7]{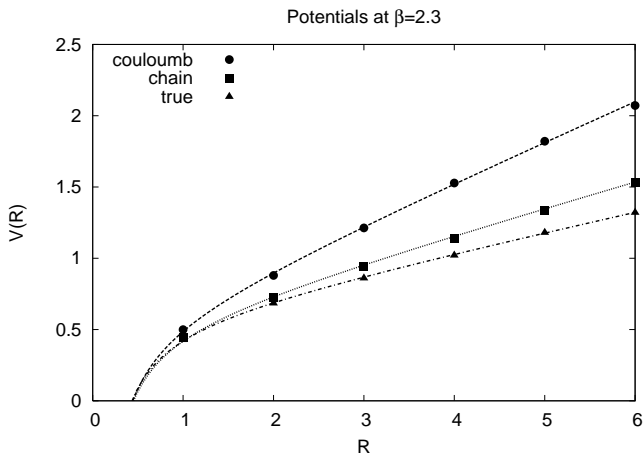}
}
\subfigure[]  
{   
 \label{pot24}
 \includegraphics[scale=0.7]{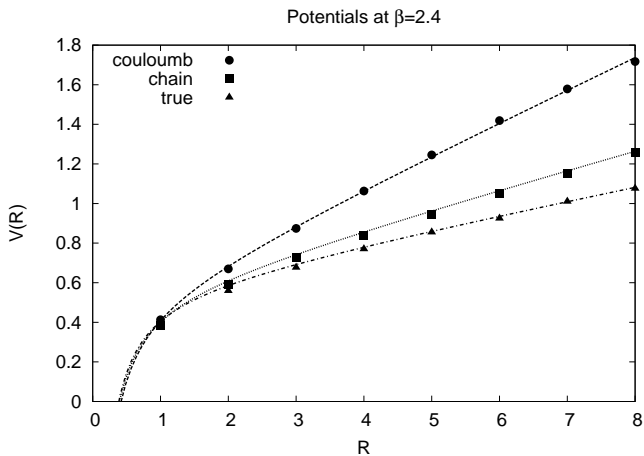}
}
\caption{The color Coulomb potential $V_{coul}(R)$, the "gluon-chain" potential $V_{chain}(R)$ derived from the variational state, 
and the static quark potential $V_{true}(R)$ extracted from ``fat-link" Wilson loops. Results are shown at lattice couplings (a) $\b=2.2$;
(b) $\b=2.3$; and (c) $\b=2.4$. Continuous lines are from a fit of data points to eq.\ \rf{fit}.}  
\label{pots}
\end{figure}

\begin{figure}[t!]
\centerline{\scalebox{0.7}{\includegraphics{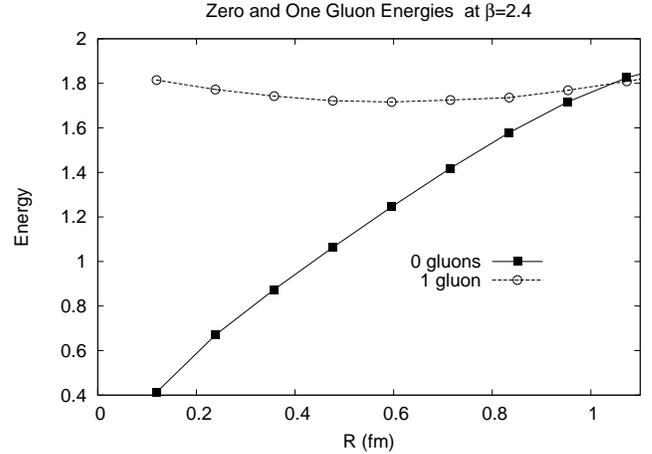}}}
\caption{Energy vs.\ quark separation of the zero-gluon and one-gluon states at $\b=2.4$.}
\label{tnn24}
\end{figure}

\subsection{Potentials}

      The results for the potential $V_{chain}(R)$ extracted from the variational state, compared to the Coulomb potential $V_{coul}(R)$, and the static quark potential $V_{true}(R)$ computed by standard methods, are shown in Fig.\ \ref{pots}.  In this and all subsequent figures, statistical errors are at most as large as the symbol sizes.  The first point to note is that inclusion of one and two gluon operators alters the slope, but not the linearity, of the potential.  The second point is that the additional one and two-gluon operators in the trial state gives an estimate for the asymptotic string tension which is much closer to the true value.  We extract string tensions, in each case, from a fit to 
\beq
            {\rm v}(R) = \s R - {\pi\over 12 R} + c
\label{fit}
\eeq
and find that while the Coulomb string tension differs from the true string tension by a factor of $2.3$ (at $\b=2.4$), the string tension of $V_{chain}(R)$ differs by roughly 38\%. The lower figure is still a significant discrepancy, but it is also a considerable improvement over the zero-gluon Coulomb result.   

    The one and two-gluon   trial-state operators, described in the previous section, are not the only ones that can be imagined, and it may be that some modest improvement in these operators would bring string tension derived from the variational state much closer to the true value.

\subsection{Gluon content}

    The energy expectation value of the one-gluon state, extracted from the one-gluon to one-gluon element of the transfer matrix, is estimated from the one-gluon to one-gluon element of the transfer matrix
\beq
             V_1(R) = - \log T_{22}
\eeq
At small quark separation, this energy greatly exceeds the Coulomb potential $V_{coul}$, so the one constituent gluon content of the minimal energy state is expected to be much smaller than the zero constituent gluon content.  However, as $R$ increases, this situation changes.  In Fig.\ \ref{tnn24}.  
we compare the energies of the zero and one-gluon states ($|\phi_1\rangle$ and optimal $|\phi_2\rangle$ states, respectively), as a function of quark separation $R$, at $\b=2.4$.  In this case $R$ is given in physical units, using the usual conversion from lattice spacing to fermis with string tension 
$\s = (440 \mbox{~Mev})^2$.   We see that the Coulomb energy (energy of the zero gluon state) rises to meet $V_1(R)$ at roughly one fermi.  It is 
reasonable to expect that the one-gluon content of the minimal energy will rise accordingly.

\begin{figure}[t!!]
\centerline{\scalebox{0.7}{\includegraphics{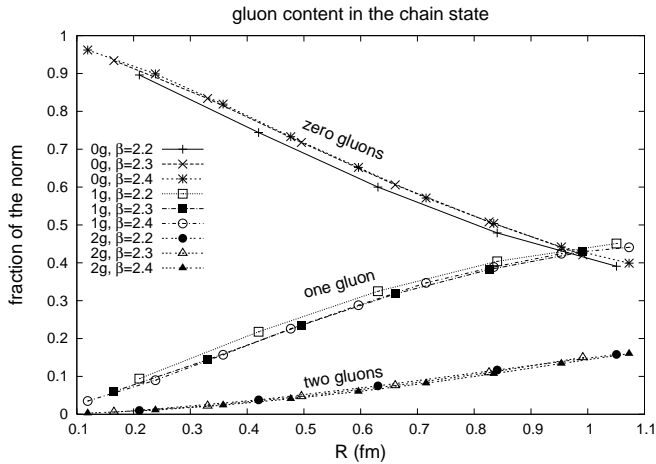}}}
\caption{Zero, one, and two-gluon content (fraction of the norm of the variational state) vs.\
quark separation $R$ in fermis, at $\b=2.2,2.3,2.4$.}
\label{gluons}
\end{figure}

      Let  
\beq
            |\psi(R)\rangle = \sum_{k=1}^6 a_k(R) |\phi_k \rangle
\eeq
denote the eigenstate of largest eigenvalue $\l_{max}$ of the $6\times 6$ transfer matrix $T_{ij}$ ,in the basis $\{|\phi_i\rangle\}$.   This is the minimal energy variational state.   By zero and one-gluon ``content" of this state, we mean the squared overlap $|\langle \phi_i |\psi\rangle |^2$ for
$i=1,2$ respectively.  Thus $a_1^2$ gives the fraction of the norm due to the zero-gluon state, $a_2^2$ the fraction due to the one-gluon state, and 
$1-a_1^2-a_2^2$ is the fraction due to the two-gluon states, orthogonal to $\phi_{1,2}$.  We have plotted these fractions vs.\ $R$ in physical units, for data at $\b=2.2,2.3,2.4$, with the result is displayed in Fig.\ \ref{gluons}.  The gluon content vs.\ $R$ in physical units is almost coupling independent.
This scaling is gratifying, and a serves as a check of the whole procedure.  It should be noted that the one-gluon content of the minimal energy state rises to equal the zero-gluon content at about one fermi, which is also the distance where the energies of the zero and and one-gluon states are roughly equal.
 
\subsection{Coulomb energy and finite size effects}

\begin{figure}[t!]
\centerline{\scalebox{0.7}{\includegraphics{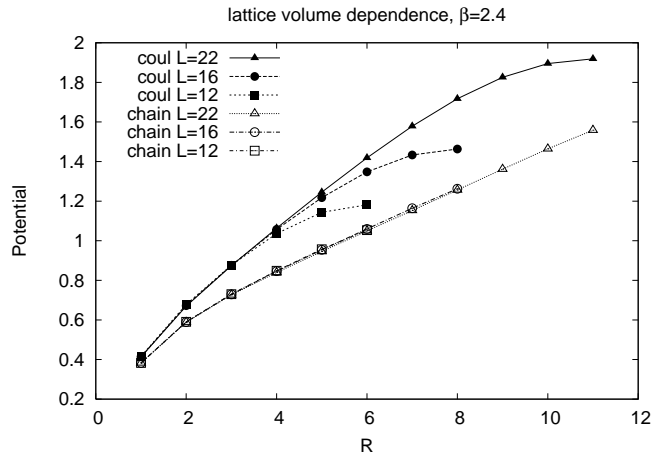}}}
\caption{Sensitivity of the Coulomb potential $V_{coul}(R)$ (solid symbols), and insensitivity of the chain potential $V_{chain}(R)$ (open symbols), to lattice volume.  Data is for the gauge coupling $\b=2.4$, and lattice volumes $L^4=12^4,16^4,22^4$.  Quark-antiquark separation $R$ is in lattice units.}
\label{coul24}
\end{figure}

    As mentioned in the Introduction, a strong objection to effective one-gluon exchange models of confinement is that such models lead inevitably to long-range dipole interactions.  We may see a hint of this ``dipole problem" in examining the lattice volume dependence of the Coulomb potential.  
In Fig.\ \ref{coul24} we plot $V_{coul}(R)$ and $V_{chain}(R)$ vs.\ $R$ in lattice units, for the coupling $\b=2.4$ and lattice volumes $12^4,16^4,22^4$.   
On all three lattice $L^4$ lattice volumes, we see that the color Coulomb potential seems to bend away from linear at $R \approx L/2$.  This departure from linearity is clearly a finite size effect, and is only seen in the vicinity of the largest possible on-axis quark separations.\footnote{Our previous work on the color Coulomb potential \cite{GO} made use of off-axis separations for $R \approx L/2$, which alleviated the distortion due to finite size effects.}  However, it is very interesting to compare the sensitivity of the color Coulomb potential to finite-size effects, with the absence of such effects in the data for $V_{chain}(R)$, which is also shown in Fig.\ \ref{coul24}.  
    
    The different sensitivities of the Coulomb and chain potentials to finite lattice size could well be associated with the dipole problem.   In particular,  the color Coulomb field associated with the zero-gluon quark-antiquark state $|\phi_1\rangle$ is not expected to be collimated in a flux tube. If the color-electric Coulomb dipole field extends throughout the lattice volume for the largest quark separations, then it is reasonable to expect a corresponding sensitivity to the finite lattice volume.  A cutoff in lattice volume would cut out a non-negligible part of the long-range field, and hence a portion of the Coulomb energy.  By contrast, if the color-electric field of the minimal energy state $|\psi\rangle$ is largely collimated in a region whose transverse dimensions are small compared to the lattice size, then even for $R \approx L/2$ one would expect much less sensitivity to the finite lattice volume.  In fact, in Fig.\ \ref{coul24}, we see no finite-size sensitivity whatever in $V_{chain}(R)$, and the chain potential remains linear out to the largest on-axis separations at each lattice volume.   This suggests that the chain state has no long-range dipole field, or at least that the 
long-range field is greatly suppressed relative to the color dipole field of the zero-gluon state.
   
\section{\label{conc}Conclusions}
 
     The color Coulomb potential is the energy of a zero constituent-gluon state in Coulomb gauge, with two static quark-antiquark charges.  The potential of such a state is known to rise linearly, at large quark separations, but its slope is far higher than the known asymptotic string tension.   What we have found in this article is that the inclusion of a few gluon constituents, in simple trial states, does not alter the linear rise of energy with quark separation, but does bring that energy very much closer to that of the true static quark potential.  According to our data, at one fermi we are only beginning to see the formation of a gluon chain state.  At that distance the energies of the zero and one-constituent gluon states are about equal, and  the zero and one-gluon states contribute equally to the minimal energy state containing a static quark-antiquark pair.   
     
     Many questions remain.  First, to what extent is the long-range dipole problem eliminated by inclusion of a few constituent gluons?  We  have seen an indication that the dipole problem is greatly reduced in the gluon-chain state, as compared to the zero-gluon state, but this needs further confirmation.    Secondly, since there is nothing compelling about the choice of operators $Q_k$ in section 3, is it possible to construct a different set of operators and corresponding basis states $\{|\phi_i\rangle\}$, such that the energy of the gluon-chain state comes significantly closer to the static quark potential?   Finally, it would be interesting to investigate chain formation for quark-antiquark sources in higher representations, and in larger gauge groups.  For example, for N-ality $k=2$ sources, do the quarks pull out a single chain of gluons, or two separate chains, as they move apart?  Are the results consistent with, e.g., Casimir scaling, or some other rule?  We hope to return to these issues in a later publication.

\acknowledgments{J.G.\ thanks Adam Szczepaniak for helpful discussions.  This research was supported in part by the U.S.\ Department of Energy under Grant No.\ DE-FG03-92ER40711 (J.G.), and by the Slovak Grant Agency for Science, Project VEGA No. 2/0070/09 (\v{S}.O.)}

\end{document}